\documentclass[prl,twocolumn]{revtex4}

\usepackage{graphicx}
\usepackage{dcolumn}
\usepackage{bm}
\usepackage{amsfonts}
\usepackage{epsfig}

\begin{document}

\title{Sub-critical statistics in rupture of fibrous materials : experiments and model.}
\author{St\'ephane Santucci, Lo\"{\i}c Vanel}
\author{Sergio Ciliberto}

\affiliation{%
Laboratoire de physique, CNRS UMR 5672, Ecole Normale Sup\'erieure
de Lyon, 46 all\'ee d'Italie, 69364 LYON Cedex 07, France
}%

\date{\today}

\begin{abstract}
We study experimentally the slow growth of a single crack in a
fibrous material and observe stepwise growth dynamics. We model
the material as a lattice where the crack is pinned by elastic
traps and grows due to thermally activated stress fluctuations. In
agreement with experimental data we find that the distribution of
step sizes follows sub-critical point statistics with a power law
(exponent 3/2) and a stress-dependent exponential cut-off
diverging at the critical rupture threshold.
\end{abstract}

\pacs{Valid PACS appear here}
\maketitle

Understanding fracture in solid materials is paramount for a safe
engineering design of structures and many efforts are still needed
to obtain a better physical picture. A puzzling observation is the
slow rupture of a material when loaded with a \emph{constant
external stress} below a critical threshold. Then, the delay time
before rupture (or lifetime of the material) strongly depends on
the applied stress. Thermodynamics has slowly emerged as a
possible framework to describe slow rupture since early
experiments have shown temperature dependence of lifetime with an
Arrhenius law \cite{Brenner62, Zhurkov65}. Statistical physics
models assuming perfect elasticity have recently proposed several
predictions for lifetime
\cite{Golubovic91,Pomeau92,Sethna9697,Kitamura97,Roux00,Scorretti01}
as well as for the average dynamics of a slowly growing crack
\cite{Santucci03}. Efforts are also made to describe slow rupture
dynamics from rheological properties of the material such as
viscoelasticity and plasticity \cite{Langer93,Chudnovsky99}.

To be able to distinguish between various theoretical
descriptions, more experimental work is needed. We present in this
letter an experiment on slow crack growth in a fibrous material.
We have observed that the crack grows by steps of various sizes
whose distribution is rather complex and evolves as a function of
the crack speed. This behavior can be explained modelling the
material as an elastic square lattice where the crack is pinned by
elastic traps and adapting the model presented in
\cite{Santucci03} to describe thermally activated growth of the
crack in an energy landscape with multiple metastable states. The
model predicts statistical distribution of step sizes in very
reasonable agreement with the experiments and has the typical
functional form obtained for sub-critical point statistics. We
stress that the material heterogeneity appears in the model only
as a characteristic mesoscopic length scale. The effect of
disorder in the material properties and the rheological behavior
have not been explicitly included in this simple model.

The experimental system we consider is a two-dimensional sheet
with a macroscopic initial crack submitted to a constant load.
This geometry is very useful to follow the crack advance using
direct observation while this would be difficult in a
three-dimensional geometry because a roughening instability of the
crack front line usually occurs. We have used a sheet of fax paper
(width $w=21cm$, length $24cm$, thickness $e=50\mu m$) for which a
natural mesoscopic length scale is the fiber size. Scanning
electron microscopy has revealed a size distribution of fibers
between $4$ and $50 \mu m$, with an average $20 \mu m$. In order
to obtain reproducible results, the fax paper was kept in a
controlled low level humidity at least one day at $\simeq 10\%$,
and also during the experiment at $5\%$. In these conditions, the
paper Young modulus is $Y=3.5GPa$. The paper sheet is mounted on a
tensile machine with both ends attached with glue tape and rolled
several times over rigid tubes. The crack is initiated at the
center of the sheet using a calibrated blade. The force $F$
applied to the sample by the tensile machine is measured by a
force gage and is perpendicular to the crack direction which
corresponds to a crack opening in a mode I configuration. During
an experiment, the crack grows and a feedback mechanism keeps $F$
constant with a precision $0.1$ to $0.5N$ and a typical time
response $10ms$. As a consequence, the stress amplitude at the
crack tip increases due to stress concentration effects and the
motion of the crack accelerates. A high resolution and high speed
digital camera (Photron Ultima 1024) is used to follow the crack
growth. Image analysis is performed to extract the length of the
crack projected on the main direction of propagation. Although the
crack actually follows a sinuous trajectory, its projected length
gives the main contribution to the stress intensity factor which
measures the amplitude of stress divergence near the tip and
verifies: $K\propto\sigma\sqrt{\ell}$, with $\sigma$ the external
constant stress applied to the sheet and $\ell$ the projected
crack length. The stress $\sigma$ is estimated from $F$ and the
area $A$ of a cross-section of the sheet, $A$ being
approximatively constant: $\sigma=F/A$. Due to the small thickness
of the paper a slight buckling occurs but it has been shown that
the scaling with stress and crack length is not significantly
modified \cite{Riks92}. On the other hand, finite width
corrections on stress intensity factor have been taken into
account: $K=g(\ell/w)\sigma\sqrt{\ell}$ with
$g(\ell/w)=[(w/\ell)\tan(\pi\ell/2w)]^{1/2}$ \cite{Lawn}.

\begin{figure}[hb]
    \centerline{\includegraphics[width=3in]{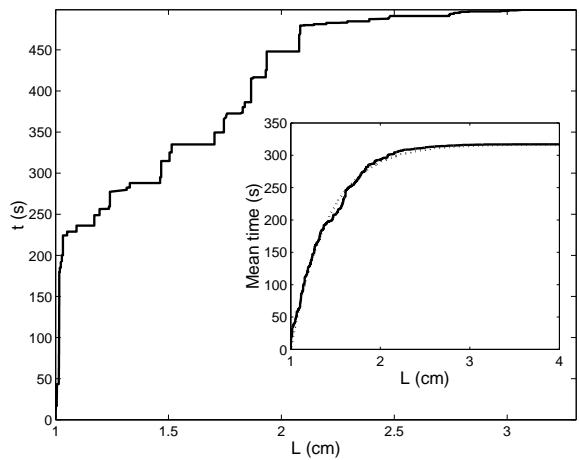}}
    \caption{Time versus crack length for a single experiment showing  crack jumps and crack arrest.
    In insert, average time to reach $L$ ($10$ experiments with $F=270\text{N}$ and
    $L_i=1\text{cm}$). The dotted curve is obtained from integration of eq.(\ref{speed1}) \cite{Santucci03}.}
    \label{growth}
\end{figure}

A typical growth curve is shown on Fig.\ref{growth}. It clearly
appears that the crack does not grow smoothly: essentially, there
are periods of rest where the crack tip does not move and periods
where it suddenly opens and advances of a certain step size $s$.
We have extensively studied the growth varying the initial crack
length ($1cm<\ell_i<4cm$) and the applied force ($140N<F<280 N$),
equivalent to an initial stress intensity factor $K_i$ between
$2.7 \text{MPa.m}^{1/2}$ and $4.2\text{MPa.m}^{1/2}$. The
resulting measured lifetime varied from a few seconds to a few
days depending on the value of the applied stress or the
temperature. Even for the same experimental conditions (same
stress, initial crack length and temperature) a strong dispersion
in lifetime was observed as expected in a model of thermally
activated growth \cite{Santucci03}. Furthermore, the average
growth dynamics shows an exponential approach of lifetime in good
agreement with the model (see insert) \cite{Santucci03}. Results
from the average dynamics will be detailed elsewhere. Here, we
want to study more extensively the step size statistics.

\begin{figure}[hb]
    \centerline{\includegraphics[width=3in]{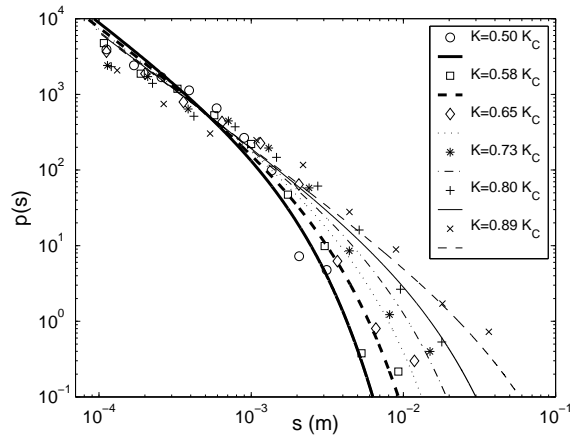}}
    \caption{Probability distribution of step sizes for various values of stress intensity factor.
    Choosing $\lambda=50 \mu\text{m}$, the different curves are the best fits of eq.\ref{distri} giving an average value
    $V=5\pm 1\text{ }\text{\AA}^3$.}
    \label{steps}
\end{figure}

It is commonly observed that the crack velocity is an increasing
function of the stress intensity factor $K$. Thus, it is natural
to look at the step statistics for a given value of $K$. In
practice, the step size distributions have been obtained for
various ranges of $K$. Fig.\ref{steps} shows the step size
distributions determined from all the data we have collected using
a logarithmic binning. Typically, $700$ data points are used to
obtain each distribution. Two regimes are observed. For small step
sizes, the distribution does not depend on the value of K, while
for larger step sizes there is a cut-off size increasing with $K$.
In practice, the toughness of the material, i.e. its critical
stress intensity factor $K_c=6.5\pm 0.05\text{MPa.m}^{1/2}$, has
been obtained as the value of $K$ beyond which the probability to
detect a jump vanishes.

\begin{figure}[hb]
    \centerline{\includegraphics[width=3in]{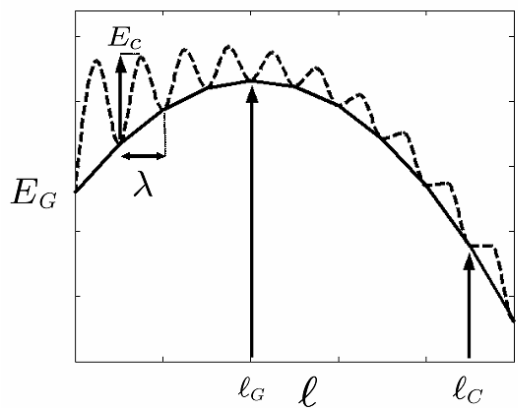}}
    \caption{Sketch of the Griffith potential energy $E_G$ as a function of crack length $\ell$ with constant applied stress (solid line).
    The energy barriers $E_C$ and the discretization scale $\lambda$ are represented by the dashed curve.}
    \label{dino}
\end{figure}

The behavior observed for the step size distributions can be
predicted using minimal physical properties. Let us assume that
the material is mainly elastic but that there is a scale at which
the material becomes discontinuous. In a perfect cristal, the only
such scale would be the atomic scale but in a fibrous material
like paper, we have an intermediate mesoscopic scale, the typical
fiber size. The elastic description of a material at a discrete
level leads to a lattice trapping effect \cite{Thomson86} with an
energy barrier which has been estimated analytically
\cite{Marder96}. To get a physical picture of the trapping in our
geometry, we have modelled numerically a 2D square lattice of
linear springs where the crack corresponds to a given number of
adjacent broken springs as described in \cite{Santucci03}. The
lattice is loaded with a constant force and we estimate the
minimum increase in potential energy needed to bring the first
spring at the crack tip at the breaking threshold. This energy is
obtained by applying an external force on the spring at the crack
tip and computing the change in elastic energy of the whole
lattice as well as the work done by the constant force at the
boundaries. We find an energy barrier per unit volume:
$E_c\simeq(\sigma_c-\sigma_m)^2/2 Y$, where $Y$ is the Young
modulus, $\sigma_c$ the material stress threshold for rupture and
$\sigma_m(<\sigma_c)$ the equilibrium local stress at the crack
tip estimated at the discrete scale $\lambda$, i.e. $\sigma_m = K
/\sqrt{\lambda}$. To each position of the crack tip corresponds a
different value of the energy barrier since the stress at the tip
increases with the crack length. Once the spring breaks, the crack
moves by at least one lattice spacing $\lambda$. The equilibrium
potential energy of the whole system is given by the Griffith
energy per unit thickness of the sheet \cite{Griffith20}:
$E_G=E_0-\pi\ell^2\sigma^2/4Y+2\gamma\ell$, where $\gamma$ is the
surface energy. On Fig.\ref{dino} we schematically represent the
energy barrier of trapping and the Griffith energy. In agreement
with previous analysis \cite{Marder96}, we find that the crack
length $\ell_c$ at which the energy barrier becomes zero is about
twice the Griffith length $\ell_G$ where the equilibrium potential
energy reaches its maximal value.

In order to model crack rupture as a thermally activated process,
we recall first ideas that were presented in \cite{Santucci03}.
Due to thermal noise at finite temperature $T$ in a fixed volume
$V$, there are statistical stress fluctuations $\sigma_f$ around
the equilibrium value $\sigma_m$ with a gaussian distribution:
$p(\sigma_f)\propto\exp[-(\sigma_f-\sigma_m)^2 V/2Y k_B T]$. The
material will break if the stress fluctuation $\sigma_f$ becomes
larger than the threshold $\sigma_c$ with a probability :
$P(\sigma_f>\sigma_c)=\int_{\sigma_c}^{\infty}
p(\sigma_f)d\sigma_f$. Assuming the rupture process is
irreversible, the velocity $v$ of the crack tip is set
proportional to the probability $P(\sigma_f>\sigma_c)$ which
gives:
\begin{equation}\label{speed1}
v=\frac{\lambda}{\tau_0}\int_{U_c}^\infty
\frac{e^{-U_f}dU_f}{\sqrt{\pi U_f}}
\end{equation}
where  $U_f=(\sigma_f-\sigma_m)^2 V/2Y k_B T$,
$U_c=U_f(\sigma_f=\sigma_c)$ and $\tau_0$ is an elementary time
scale (typically, an inverse vibrational frequency). Integration
of eq.(\ref{speed1}) gives the average growth curve in insert of
Fig.\ref{growth}.

We extend now this model to describe thermally activated and
irreversible motion of a crack in the rugged potential energy
landscape introduced above. Below $\ell_c$, the energy barriers
$E_c(\sigma_m)$ trap the crack in a metastable state for an
average time $\tau_p$ depending on the barrier height.
Irreversible crack growth is a very reasonable assumption when
$\ell>\ell_G$ since the decrease in equilibrium potential energy
makes more likely for the crack to open than to close. When a
fluctuation $\sigma_f$ occurs, it will increase locally the free
energy per unit volume by
$E_f(\sigma_m)\simeq(\sigma_f-\sigma_m)^2/2Y$ (this comes from a
Taylor expansion of the bulk elastic free energy in agreement with
the numerical estimate of $E_c$ and the gaussian form of the
stress fluctuations). The energy $E_f$ can be used by the crack to
overcome the barrier. If there are no dissipative mechanisms the
crack will grow indefinitely when $\ell>\ell_G$ as the barriers
get smaller and smaller and the release of elastic energy helps to
reach a more energetically favorable position. We introduce a
simple mechanism of crack arrest assuming that after overcoming
the energy barrier the crack looses an energy identical to the
barrier size and does not gain any momentum from the elastic
release of energy (experimentally, dissipation will come from
acoustic wave emissions, viscous or plastic flow, etc.). When the
crack reaches the next trap it still has an energy $E_f-E_c$ which
might be sufficient to overcome the next barrier. For a given
fluctuation energy $E_f$, the crack will typically have enough
energy to overcome a number of barriers
$n=E_f(\sigma_m)/E_c(\sigma_m)$ and make a jump of size $s=n
\lambda$ (the decrease of $E_c(\sigma_m)$ with $\sigma_m$ during a
jump of size $s$ has been neglected). The probability distribution
for $E_f$ is explored at each elementary step $\tau_0$, while the
probability distribution of step size is explored after each
average time $\tau_p$ spent in the trap. In order to relate the
two probabilities, we express the mean velocity in a different way
as the ratio of the average step size to the average trapping
time:
\begin{equation}\label{speed2}
v=\frac{\int_\lambda^\infty s p(s) ds}{\tau_p}
\end{equation}
From the identity between eq.(\ref{speed1}) and eq.(\ref{speed2}),
and the normalization condition of the probability
($\int_{\lambda}^\infty p(s)ds=1$), we obtain the probability
distribution :
\begin{equation}\label{distri}
p(s)=N(U_c)\frac{\sqrt{\lambda}e^{-s/\xi}}{2 s^{3/2}}
\end{equation}
 where $N(U_c)=\left[e^{-U_c}-\sqrt{\pi
U_c}\text{erfc}(\sqrt{U_c})\right]^{-1}$ and $\xi=\lambda/U_c$. We
find a power law with an exponent $3/2$ and an exponential cut-off
with a characteristic length $\xi\sim(\sigma_c-\sigma_m)^{-2}$
diverging at the critical stress $\sigma_c$. Incidently, we note
that this probability has a form similar to sub-critical point
probability distributions in percolation theory \cite{Stauffer91}.
From eq.(\ref{distri}), we can compute from this distribution the
average and variance of step sizes:

\begin{equation}\label{mean}
\langle s \rangle=N(U_c)\frac{\lambda\sqrt{\pi}}{2
\sqrt{U_c}}\text{erfc}(\sqrt{U_c})
\end{equation}

\begin{equation}\label{var}
\langle s^2 \rangle=N(U_c)\frac{\lambda^2\sqrt{\pi}}{4
U_c^{3/2}}\left(\text{erfc}(\sqrt{U_c})+2
\sqrt{\frac{U_c}{\pi}}e^{-U_c}\right)
\end{equation}

We obtain two asymptotical behaviors. When the relative energy
barrier is high ($U_c\gg 1$), $\langle s \rangle\simeq\lambda$ and
$\langle s^2 \rangle\simeq\lambda^2$. In this limit, there is only
one step size possible. When the relative energy barrier becomes
low ($U_c\ll 1$), we predict a divergence at critical point :
$\langle s \rangle\sim(\sigma_c-\sigma_m)^{-1}$ and $\langle s^2
\rangle\sim(\sigma_c-\sigma_m)^{-3}$. Then, the crack velocity is
expected to be dominated by the critical divergence of crack
jumps.

\begin{figure}[hbt]
    \centerline{\includegraphics[width=3in]{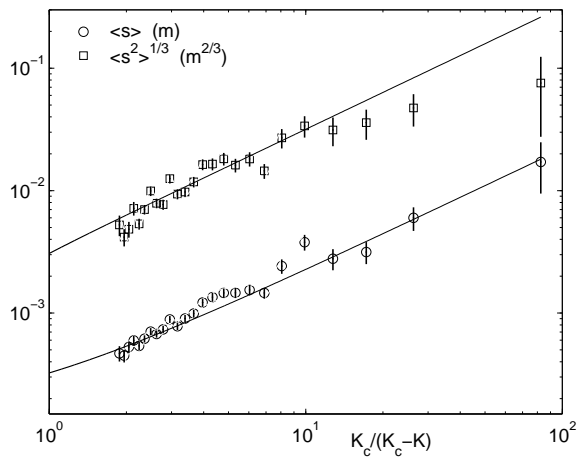}}
    \caption{The mean and cubic root of the variance from raw measurements of step sizes is well reproduced by
     the model (eq.(\ref{mean}) and eq.(\ref{var})) plotted with $\lambda=50\mu\text{m}$ and
    $V=5\text{\AA}^3$. }
    \label{moments}
\end{figure}

To compare with our experimental data, we use an estimate of the
stress near the crack tip by assuming as above that
$\sigma_m=K/\sqrt{\lambda}$. In addition, the normalization
condition of the distribution actually reduces the model to one
parameter, the ratio $V/\lambda^2$. In our model, $\lambda$
represents the mesoscopic scale of discretization in paper.
Setting $\lambda=50\mu\text{m}$, $V$ is the only unknown. One
parameter fits of step size distributions in Fig.\ref{steps} for
each range of stress intensity factors give very robust results:
$V=5\pm 1\text{ }\AA^3$. To check the asymptotic limit close to
the critical point, we have plotted in Fig.\ref{moments} $\langle
s \rangle$ and $\langle s^2 \rangle^{1/3}$ as a function of
$K_c/(K_c-K_m)$. Here, the mean and the variance of step sizes
have been computed from the raw measurements in a given range of
$K$. Because it requires less statistics to estimate the first two
moments of the distribution than the distribution itself, we are
able to narrow the width of the $K$
range for each data point without changing the global trend. The
solid lines represents the model prediction using the fitted value
of $V$ from the distributions of Fig.\ref{steps}. Not only the
model reproduces reasonably well the evolution of the step size
distributions with $\sigma_m$ ($V$ is essentially constant and all
the other parameters are fixed), but the asymptotic divergence of
the first two moments of the distribution are also well
reproduced. For the mean step size, the scaling is observed up to
$K$ values very close to $K_c$ ($1\%$). In the model, we see that
the ratio of the standard deviation of the distribution over the
mean size is diverging at $K_c$. Thus, close to $K_c$, the measure
of variance becomes more inaccurate than the measure of the mean.

The value obtained for the volume $V$ is at the atomic scale
($V^{1/3}\simeq 1.7\AA$). Its small value gives an idea of the
microscopic scale at which the thermodynamical stress fluctuations
have the proper amplitude to trigger rupture in our model. It
should be realized that the model actually predicts a lower limit
for this microscopic scale. First, it assumes a strong dissipation
of energy during crack advance since none of the elastic release
of energy is used to keep the crack moving. This is certainly an
overestimation of a real dissipative mechanism, would it be
viscoelastic or plastic. Decreasing dissipation in the model will
permit larger steps of the crack. In order to obtain the same
experimental velocity, the trapping time must also be larger which
will happen if the rupture occurs at a larger microscopic scale.
Second, disorder in the material properties has been completely
neglected. It has been shown recently that disorder effectively
reduce the energy cost for breaking and this will also permit
rupture at a larger microscopic scale \cite{Scorretti01}. Further
theoretical work needs to be done to introduce a more realistic
dissipative mechanism and take into account disorder in the
material properties. As the model stands now, we believe that it
should apply to any elastic materials for which a structure at a
mesoscopic scale exists. For example, it would be interesting to
understand if the model can explain crack jump dynamics observed
in semi-cristalline polymers \cite{Parsons99}. To conclude, we
have shown that a simple model of thermally activated crack
dynamics is able to reproduce with a good accuracy the step size
distribution of the crack growth. This is quite interesting
because it may open new perspectives in the description of rupture
as a thermally activated process.

 \acknowledgments
 We acknowledge illuminating discussions with E. Bouchaud, J.-P. Bouchaud, M. Ciccotti, and M.
 Marder.

\end{document}